\documentclass[12pt,letterpaper]{article}

\parskip=\medskipamount            
\def\7#1#2{\mathop{\null#2}\limits^{#1}}        

\def\beee{\begin{equation}}
\def\eeee{\end{equation}}
\def\dggg{^{\dagger}}

\def\inot{i \!\not \! \partial}
\def\nott{\not \!}
\def\nottt{\not \!\!}
\begin{document}

\bibliographystyle{unsrt}
\begin{center}
\textbf{HYBRID DIRAC FIELDS}\\
[5mm]
O.W. Greenberg\footnote{email address, owgreen@physics.umd.edu.}\\
{\it Center for Theoretical Physics\\
Department of Physics \\
University of Maryland\\
College Park, MD~~20742-4111}\\
University of Maryland Preprint PP-03043\\
~\\

\end{center}

\begin{abstract}

Hybrid Dirac fields are fields that are general superpositions of the
annihilation and creation parts of
four Dirac spin 1/2 fields, $\psi^{(\pm)}(x;\pm m)$, 
whose annihilation and creation parts obey the Dirac
equation with mass $m$ and mass
$-m$. We discuss a specific case of such fields, which has been called
``homeotic.'' We show for this case,
as is true in general for hybrid Dirac fields (except the ordinary fields
whose annihilation and creation parts both obey one or the other Dirac
equation), that (1) any interacting
theory violates both Lorentz covariance and causality,
(2) the discrete transformations $\mathcal{C}$, and 
$\mathcal{CPT}$ map the pair $\psi_h(x)$ and 
$\bar{\psi}_h(x)$ into fields that are not linear
combinations of this pair, and (3) the chiral projections of $\psi_h(x)$
are sums of the usual Dirac fields with masses $m$ and $-m$; on 
these chiral projections $\mathcal{C}$, and $\mathcal{CPT}$ are defined
in the usual way, their interactions do not violate $\mathcal{CPT}$,
and interactions of chiral projections are Lorentz covariant and causal. 
In short, the main claims concerning ``homeotic'' fields are incorrect.

\end{abstract}

1. Introduction

Since a spin 1/2 field on which parity is defined can obey either 
of two Dirac equations
\beee
(i \!\not \! \partial -m)\psi(x;m)=0                   \label{m}
\eeee
or
\beee
(i \!\not \! \partial +m)\psi(x;-m)=0                         \label{-m}
\eeee
and since the positive (annihilation) and negative (creation) 
frequencies
of a free field can be separated in a Lorentz covariant way, one 
can consider
a family of free ``hybrid Dirac fields'' which, with suitable 
normalizations, are
linear combinations of the annihilation and creation parts of 
mass $m$
and mass $-m$ Dirac fields. This type of field, with a specific 
choice given below, was considered in
a recent paper by G. Barenboim and J. Lykken~\cite{bar} (BL) in 
connection with a proposed model of
$\mathcal{CPT}$ violation for neutrinos. They called their field
a ``homeotic'' field. The purpose of this paper is to study the 
properties of this type of 
field which, for reasons stated above, I prefer to call a ``hybrid 
Dirac field.'' 
BL proposed their ``homeotic'' field as a counter-example to my 
general theorem~\cite{gre} that interacting fields that violate $\mathcal{CPT}$
symmetry necessarily violate Lorentz covariance. I will show below 
that the BL example does not violate $\mathcal{CPT}$ and that their
interacting ``homeotic'' field violates both Lorentz covariance 
and causality.

The simplest representation of the BL field is
\beee
\psi_h(x)= \psi^{(+)}(x;m) + \psi^{(-)}(x;-m),
\eeee
where to be explicit and to establish notation,
\beee 
\psi^{(+)}(x;\pm m)=\frac{1}{(2 \pi)^{3/2}}\int
\frac{d^3p}{2E_p} \sum_{\pm s} b(p,s)u(p,s;\pm m) exp(-i p \cdot x), 
\eeee 
\beee 
\psi^{(-)}(x;\pm m)=\frac{1}{(2 \pi)^{3/2}}\int
\frac{d^3p}{2E_p}\sum_{\pm s} d\dggg(p,s)v(p,s;\pm m) 
exp(i p \cdot x), 
\eeee
\beee 
\bar{\psi}^{(-)}(x;\pm m)=\frac{1}{(2 \pi)^{3/2}}\int
\frac{d^3p}{2E_p} \sum_{\pm s} b\dggg(p,s)\bar{u}(p,s;\pm m) 
exp(ip\cdot x), 
\eeee 
\beee
\bar{\psi}^{(+)}(x;\pm m)=\frac{1}{(2 \pi)^{3/2}}\int
\frac{d^3p}{2E_p}\sum_{\pm s} d(p,s)\bar{v}(p,s;\pm m) 
exp(-ip\cdot x).
\eeee
Note the difference between $\overline{\psi^{(+)}}=\bar{\psi}^{(-)}$ and 
$\bar{\psi}^{(+)}$, etc. To separate $\psi_h(x)$ into $\psi^{(+)}(x;m)$
and $\psi^{(-)}(x;-m)$, use 
$[(\pm \inot_x + m)/2m]\psi_h(x)=\psi^{(\pm)}(x;\pm m)$; $b(p,s)$, etc., can
then be calculated in the usual way.
The annihilation and creation operators are normalized covariantly,
\beee
[b(p,s), b\dggg(q,t)]_+=2E_p\delta(\mathbf{p}-\mathbf{q})\delta_{st},
\eeee
etc., and the spinors obey
\beee 
(\nott p \mp m) u(p,s;\pm m) = 0, 
\eeee 
\beee 
(\nott p \pm m) v(p,s;\pm m) = 0,
\eeee
with normalizations 
$\bar{u}(p,s;\pm m)u(p,t;\pm m)=\pm 2m\delta_{st}$
and $\bar{v}(p,s;\pm m)v(p,t;\pm m)=\mp 2m\delta_{st}$. 
I chose these normalizations so that going from 
$\psi^{(\pm)}(x;\pm m)$ to $\psi^{(\pm)}(x;\mp m)$ is accomplished
by changing the sign of $m$. (This would not be the case if I had
chosen $\bar{u} u=\delta_{st}$, for example.) I use the same 
non-chiral Dirac matrices of~\cite{bjo} and~\cite{itz} and the same
creation and annihilation operators for both the $m$ and $-m$ 
fields. With 
this choice of gamma matrices, the $m$  $u$ spinors have large upper two 
components and the $-m$ $u$ spinors have large lower two components, and 
vice versa for the $v$ spinors. The reversal of the signs of the normalizations
of the $u$ and $v$ spinors for the $-m$ case may seem strange; however these
normalizations are consistent with the anticommutation relation for the
$-m$ field, with the positivity of energy (note that the $u\dggg u$ 
normalization is positive for both the $m$ and $-m$ spinors) and with the
relation between $m$ and $-m$ fields via $\gamma^5$ given below. If 
I had
started with the $-m$ field rather than with the $m$ field, 
I could have
chosen gammas so that the $-m$ $u$ spinors would have positive 
normalization.

(Other linear combinations of the positive and negative frequency 
parts of 
Dirac fields of mass $m$ and mass $-m$ also fall into the category 
of hybrid
Dirac fields as do the usual Dirac fields, but in this paper I 
discuss only
the example given above). I follow the universal convention,
~\cite{bjo},~\cite{itz} and~\cite{pes}, that the 
$p^0=E_{\mathbf{p}}=\sqrt{m^2+\mathbf{p}^2}$ in
spinors is always the positive energy; unfortunately BL
violate this convention.

To see that my representation of $\psi_h(x)$ is identical with the BL field,
calculate the anticommutator
\beee
[\psi_{h\alpha}(x),\bar{\psi}_{h\beta}(y)]_+=
(\inot_x +m)_{\alpha \beta}\Delta^{(+)}(x-y)
-(\inot_x -m)_{\alpha \beta}\Delta^{(-)}(x-y),
\eeee
where
\beee
\Delta^{(+)}(x)=\frac{1}{(2\pi)^3}\int \frac{d^3p}{2E_p}
exp(-i p \cdot x),~~\Delta^{(-)}(x)=\Delta^{(+)}(-x). 
\eeee
This agrees with
(2.14) of BL, (their $D(x)=\Delta^{(+)}(x)$), except that their Dirac 
indices are transposed; however, 
this differs from the Dirac result by the sign of the second term
proportional to $m$, not that of the first term, as stated by BL.
It is instructive to rewrite this result as
\beee
[\psi_{h\alpha}(x),\bar{\psi}_{h\beta}(y)]_+=
(\inot_x)_{\alpha \beta}i\Delta(x-y)+m\delta_{\alpha \beta}\Delta^{(1)}(x-y),
\eeee
where $i \Delta(x) =\Delta^{(+)} -\Delta^{(-)}$ and
$\Delta^{(1)}=\Delta^{(+)}+\Delta^{(-)}$, since this separates the local and
nonlocal terms.

In addition,
$\inot \psi_h(x)=m(\psi^{(+)}(x;m)-\psi^{(-)}(x;-m))$;
it is straightforward to check that
\beee
m(\psi^{(+)}(x;m)-\psi^{(-)}(x;-m))=\frac{-im}{\pi}\mathbf{P}\int dt^{\prime}
\frac{1}{x^0-t^{\prime}}\psi_h(t^{\prime},\mathbf{x})
\eeee
(provided the integral is defined suitably). Thus the present
$\psi_h$ obeys the equation of motion (2.7) of BL.

In agreement with my result~\cite{gre} and with BL, free, or generalized
free, hybrid Dirac 
fields transform covariantly. This is connected with the existence  
of
$\theta(\pm p^0)$ which exists for timelike momenta, but does not 
exist for spacelike momenta. I also agree with BL that the 
propagator of their field $\psi_h(x)$ is not causal; it is also not
covariant.

From Eq.(\ref{m},\ref{-m}), $\gamma^5 \psi(x;\pm m)$ obeys the Dirac
equation for $\psi(x;\mp m)$, so 
it is useful to consider the discrete transformation
\beee
\mathcal{M} \psi(x;m) \mathcal{M}\dggg =\psi(x;-m)=\gamma^5 \psi(x;m),
\eeee
\beee
\mathcal{M} \psi(x;-m) \mathcal{M}\dggg =\psi(x;m)=\gamma^5 \psi(x;-m).
\eeee
This shows that the $m$ and $-m$ spinors are related by $\gamma^5$.

The calculation of the BL current is particularly simple for the space
components. The result is
\begin{eqnarray}
[J^i(x),J^j(y)]_-&=&
\bar{\psi}_h(x)\gamma^i \inot_x i\Delta(x-y)\gamma^j\psi_h(y)
-\bar{\psi}_h(y)\gamma^j \inot_y i\Delta(y-x)\gamma^i\psi_h(x) 
\nonumber \\
& &
+m\{-\delta^{ij}[\bar{\psi}_h(x)\psi_h(y)-\bar{\psi}_h(y)\psi_h(x)] 
\nonumber  \\
& &
-i[\bar{\psi}_h(x)\sigma^{ij}\psi_h(y)+\bar{\psi}_h(y)\sigma^{ij}\psi_h(x)]\}
\Delta^{(1)}(x-y).
\end{eqnarray}
As we expect, the term proportional to the gradient which is the same for
$m$ and for $-m$ is local, but the term proportional to $m$ is proportional
to $\Delta^{(1)}(x-y)$ and is not local. In particular, 
\beee
\Delta^{(1)}(0,\mathbf{x})=\frac{1}{(2 \pi)^3}\int \frac{d^3p}{2E_p}
cos~\mathbf{p}\cdot\mathbf{x} \neq 0, \mathbf{x} \neq 0.
\eeee
Note that for the space indices $i~,j$ the BL current is the same as
what they call the ``Dirac'' current. BL correctly point out that 
causality fails to hold for their ``Dirac'' current, their Eq.(3.11),
but do not notice that this fact directly contradicts their 
(incorrect) equal time current commutation relations for the space
indices, their Eq.(3.3).
Thus the statements that BL make concerning the causality condition 
(their Eq.(3.7))
\beee
[\mathcal{H}_I(\mathbf{x}),\mathcal{H}_I(\mathbf{y})]_- 
\propto \delta(\mathbf{x}-\mathbf{y}),
\eeee
the Lorentz covariance of the time-ordering in the Dyson series, and the Lorentz 
invariance of their $S$-matrix are all incorrect
and their interacting theory is
neither Lorentz covariant nor causal.

In contrast to my disagreement with the assertions concerning equal time
commutation relations for the currents and the Hamiltonian density, I agree
with BL about the equal time commutation relations for the chiral projections
of the hybrid fields. The agreement here is because \textit{the chiral 
projections of the hybrid fields are sums of the usual Dirac fields with masses
$m$ and $-m$. These chiral projections are not hybrid (or homeotic) fields at all!}

I agree with BL that parity and time reversal are realized in the 
usual way. 
BL assert that charge conjugation is realized in a different way
than usual. This is incorrect. This assertion seems to be based on the
tacit assumption that $\mathcal{C}$ and 
$\mathcal{CPT}$ map the pair $\psi_h(x)$ and 
$\bar{\psi}_h(x)$ onto themselves. Unexpectedly, this is not the case. 
The discrete transformations $\mathcal{C}$ and 
$\mathcal{CPT}$ map the terms in $\psi_h(x)$ into terms that are not
present in $\bar{\psi}_h(x)$ and map the terms in $\bar{\psi}_h(x)$ into terms
that are not present in $\psi_h(x)$. 

Although it is well known, I emphasize that the requirement that
charge conjugation changes the sign of the field and the requirement that
charge conjugation interchanges particle and antiparticle are 
equivalent~\cite{bjo2,itz2,wei}.
Here is the standard argument concerning charge conjugation~\cite{itz}.
Thus if $\psi(x;m)$ obeys
\beee
[\inot_x -e \not \!\! A(x) -m] \psi(x;m)=0,    \label{1}
\eeee
then the charge conjugate field $\psi^{c}(x;m)$
must obey
\beee
[\inot_x +e \nottt A(x) -m] \psi^{c}(x;m)=0;
\eeee
and if $\psi(x;-m)$ obeys
\beee
[\inot_x -e \nottt A(x) +m] \psi(x;-m)=0,
\eeee
then the charge conjugate field $\psi^{c}(x;-m)$
must obey
\beee
[\inot_x +e \nottt A(x) +m] \psi^{c}(x;-m)=0.
\eeee
For example, take the adjoint of both sides of Eq.(\ref{1}) to get
\beee
\psi^{\dagger}(x;m)[-\inot_x\dggg -e
\nottt A\dggg(x) -m]=0
\eeee
where the derivative acts to the left. 
Next multiply from the right by $\gamma^0$ to get 
\beee
\bar{\psi}(x;m)[-\inot_x -e \nottt A(x) -m]=0.
\eeee
Next transpose the equation to get
\beee
[-\inot_x^T -e \nottt A(x)^T -m]
\bar{\psi}^{T}(x;m).
\eeee
Finally multiply by the usual $C$ matrix that obeys 
$C \gamma^{\mu~T}C\dggg=-\gamma^{\mu}$ to get 
\beee
[\inot_x +e\nottt A(x) -m]C\bar{\psi}^T(x;m)= 0.
\eeee
Thus $\psi^{c}(x;m)=C \bar{\psi}^{T}(x;m)$. This relation holds 
separately for the annihilation and creation parts,
\beee
\mathcal{C} \psi^{(\pm)}(x;m) \mathcal{C}\dggg  =  \psi^{c(\pm)}(x;m) =
 C \bar{\psi}^{(\pm)T}(x;m),
\eeee
and
analogous relations hold for the relation of $\psi^{(\pm)}(x;-m)$ to
$\psi^{c(\pm)}(x;-m)$ up to a minus sign. 

Since
\beee
\psi_h(x)= \psi^{(+)}(x;m) + \psi^{(-)}(x;-m),
\eeee
charge conjugation takes $\psi_h(x)$ to 
\beee
\psi^c_h(x)=C \bar{\psi}^{(+)T}(x;m) + C \bar{\psi}^{(-)T}(x;-m).
\eeee
The Pauli adjoint of $\psi_h(x)$ is
\beee
\bar{\psi}_h(x)= \bar{\psi}^{(+)}(x;-m) +\bar{\psi}^{(-)}(x;m);
\eeee
thus neither of the terms in $\psi^{c}_h(x)$ is present in
$\bar{\psi}_h(x)$. Another way to look at this is to note that
charge conjugation takes, for example, a $b$ annihilation operator to
a $d$ annihilation operator; however, the $b$ operator is in 
$\psi^{(+)}(x;m)$, 
while the $d$ operator is in $\bar{\psi}^{(+)}(x,m)$,
therefore it takes $\psi^{(+)}(x;m)$ to 
$\bar{\psi}^{(+)}(x;m)$, which does not appear in $\bar{\psi}_h(x)$. 
A third way to see this is to note that the $b$ 
annihilation operator 
in $\psi_h$ is
associated with a $u(p,s;m)$ spinor that has large upper components and 
should be transformed into $C \bar{u}^T(p,s;m)$ which has large lower
components, while the
$d$ annihilation operator in $\bar{\psi}(x;-m)$ is associated with a 
$\bar{v}(p,s;-m)$ spinor that has
large upper components. Thus the discrete transformations $\mathcal{C}$, and 
$\mathcal{CPT}$ map the pair $\psi_h(x)$ and 
$\bar{\psi}_h(x)$ into other fields that are not linear combinations
of this pair, and the statements of
BL concerning $\mathcal{C}$ and $\mathcal{CPT}$ are incorrect.
(I have suppressed the usual phases that can accompany the definitions
of the discrete transformations. The reader who wishes can supply 
these phases; the conclusions remain unchanged.)

Independent of the discussion just given above, 
it is important to point out that $\mathcal{CPT}$ has a more basic role in
relativistic quantum field theory than any of the other discrete transformations
$\mathcal{C}$, $\mathcal{P}$, $\mathcal{T}$ or their bilinear products. 
$\mathcal{CPT}$ is the unique discrete symmetry that can be connected to the
identity when the proper orthochronous Lorentz group, $L^{\uparrow}_+$, and its
associated covering group, $SL(2,C)$, are enlarged to the proper complex
Lorentz group, $L_+(C)$,  and its covering group, $SL(2,C) \otimes SL(2,C)$.
This is not to say that Lorentz invariance alone leads to $\mathcal{CPT}$ 
symmetry. In order for Lorentz invariance to imply $\mathcal{CPT}$ symmetry
it is necessary and sufficient that a relaxed form of spacelike commutativity 
(or anticommutativity) called ``weak local commutativity'' 
should hold~\cite{jos}. This last remark shows why a \textit{free} field with different
masses for particle and antiparticle can be Lorentz invariant on-shell and yet not
obey $\mathcal{CPT}$ symmetry. The reason is that such a field does not obey
weak local commutativity. Of course the Green's functions of such a field will
not be Lorentz invariant.

In terms of the irreducible representations of $L^{\uparrow}_+$ 
\beee
\psi(x;m)=\psi_{(1/2,0)} \oplus \psi_{(0,1/2)}
\eeee
and 
\beee
\psi(x;-m)=\psi_{(1/2,0)} \ominus \psi_{(0,1/2)},
\eeee
where $(1/2, 0)$ is the representation with one undotted index and 
$(0, 1/2)$ is the representation with one dotted index in van der Waerden's
notation~\cite{van}. See also~\cite{jos,str}. 
The annihilation and creation parts of these fields each have the
corresponding decomposition in irreducibles of $SL(2,C)$.
As stated above one
can define $\mathcal{CPT}$ without ever considering the individual discrete
symmetries. 
Pauli~\cite{pau} showed that $\mathcal{CPT}$ takes each irreducible
representation of the homogeneous Lorentz group (without discrete symmetries) 
into the adjoint of the same irreducible. See also~\cite{wei2}.
Thus one can consider $\mathcal{CPT}$
acting on each of the four $\psi^{(\pm)}(x;\pm m)$ as well as on each of
their decompositions into irreducibles separately, regardless of
whether or not any of the other ones are added to it to form a hybrid 
(or homeotic) field.

Using the relation
$\psi(x;-m)=\gamma^5 \psi(x;m)$ we can induce the discrete 
transformations of the $-m$ fields from those of the $m$ fields. Thus
\beee
\mathcal{P} \psi(x;-m) \mathcal{P}\dggg =-\gamma^0 \psi(i_s x;-m),
~i_s x=(x^0,-x^i)
\eeee
\beee
\mathcal{C} \psi(x;-m) \mathcal{C}\dggg =
i \gamma^2 \psi^{\dagger T}(x;-m),
\eeee
\beee
\mathcal{T} \psi(x;-m) \mathcal{T}\dggg =
i\gamma^1 \gamma^3 \psi(i_t x;-m),~i_t x=(-x^0,x^i)
\eeee

Because $\mathcal{CPT}$ does not map hybrid Dirac fields and their
adjoints onto themselves, 
if the hybrid Dirac field is coupled linearly to a usual
Dirac field, the resulting bilinear term violates $\mathcal{CPT}$ and produces an 
interaction that violates both Lorentz covariance in the sense that its $T$-product
will not be covariant, and causality in the 
sense that it fails to commute at spacelike separation. By contrast, \textit{the 
chiral projections of hybrid Dirac fields are sums of the usual
Dirac fields with mass $m$ and $-m$}. For example,
\begin{eqnarray}
\frac{1+\gamma^5}{2} \psi^h(x) & = & \frac{1+\gamma^5}{2}
(\psi^{(+)}(x;m)+\psi^{(-)}(x;m)),
\nonumber \\
&=&  \frac{1+\gamma^5}{2}\psi(x;m)=\frac{1}{2}(\psi(x;m)+\psi(x;-m))
\end{eqnarray}
and 
\beee
\frac{1-\gamma^5}{2} \psi^h(x)=
\frac{1-\gamma^5}{2}(\psi^{(+)}(x;m)-\psi^{(-)}(x;m)).
\eeee
(The relative minus sign between $\psi^{(+)}(x;m)$ and 
$\psi^{(-)}(x;m)$ in this last equation is not significant since 
$\psi^{(+)}(x;m) \pm \psi^{(-)}(x;m)$ both have the same anticommutation
relations with their Pauli adjoints and thus have the same 
contractions, so all their observable expectation values are the same.)
$\mathcal{CPT}$ acts in the usual way on both $\psi(x;m)$ and 
$\psi^{(+)}(x;m)-\psi^{(-)}(x;m)$; thus the
bilinear terms that couple the chiral projections of $\psi_h$ 
to a chiral Dirac field preserve $\mathcal{CPT}$ and are Lorentz 
invariant and causal. This means that the terms in (4.1) of BL do not
violate $\mathcal{CPT}$ and thus their model fails as an example of 
$\mathcal{CPT}$ violation.
 
Conclusions: Although free hybrid (or ``homeotic'') Dirac fields can be 
Lorentz covariant on-shell, 
interacting ones necessarily violate Lorentz covariance in agreement with the
theorem in~\cite{gre}. Such fields also violate causality. Free chiral hybrid 
Dirac fields are sums of the usual Dirac fields with masses $m$ and $-m$ and 
because of that they can
be local and Lorentz covariant. Further, since they are sums of the usual
Dirac fields they must have the usual $\cal{CPT}$ transformation. 
This means that their bilinear coupling to a usual chiral
Dirac field does not violate $\cal{CPT}$. The
suggestion that ``acausal propagation combined with nonlocal 
interactions yield a causal theory''~\cite{bar} is incorrect. It seems unlikely
that hybrid (or ``homeotic'') Dirac fields will be of phenomenological 
importance. 

Acknowledgements: I thank Axel Krause, Markus Luty and Rabi Mohapatra for
helpful comments. Correspondence with Gabriela Barenboim and Joe Lykken has
helped me to clarify some of the issues discussed in this paper. This work
was supported in part by the National Science Foundation, 
Award No. PHY-0140301. I am particularly happy to thank Lev Okun for his
interest in this work.


\begin{thebibliography}{20}

\bibitem{bar} G. Barenboim and J. Lykken, Phys. Lett. B554, 73 (2003).

\bibitem{gre} O.W. Greenberg, Phys. Rev. Lett. 89, 231602 (2002).

\bibitem{bjo} J.D. Bjorken and S.D. Drell, \textit{Relativistic Quantum
Mechanics}, (McGraw-Hill, New York, 1964) and \textit{Relativistic
Quantum Fields}, ibid (1965).

\bibitem{itz} C. Itzykson and Zuber, \textit{Quantum Field Theory},
(McGraw-Hill, New York, 1980).

\bibitem{pes} M.E. Peskin and D.V. Schroeder, \textit{An Introduction
to Quantum Field Theory}, (Addison Wesley, New York, 1995).

\bibitem{bjo2} Bjorken and Drell, \textit{op. cit.}, pp115-117.

\bibitem{itz2} Itzykson and Zuber, \textit{op. cit.}, pp152-153.

\bibitem{wei} S. Weinberg, \textit{The Quantum Theory of Fields, Vol. 1},
(Cambridge, New York, 1995), pp241-242.

\bibitem{jos} R. Jost, Helv. Phys. Acta 30, 409 (1957).

\bibitem{pau} W. Pauli, in \textit{Niels Bohr and the Development of Physics},
(McGraw-Hill, New York, 1955), pp30-51.

\bibitem{van} B.L. van der Waerden, \textit{Group Theory and Quantum Mechanics},
(Springer, Berlin, 1974).

\bibitem{str} R.F. Streater and A.S. Wightman, \textit{PCT, Spin \& Statistics,
and All That}, (Benjamin, New York, 1964).

\bibitem{wei2} Weinberg, \textit{op. cit.}, pp 244-246.
  
\end{thebibliography}
\end{document}